# Phased-MIMO Radar: A Tradeoff Between Phased-Array and MIMO Radars

Aboulnasr Hassanien, *Member, IEEE* and Sergiy A. Vorobyov\*, *Senior Member, IEEE*

## Abstract

We propose a new technique for multiple-input multiple-output (MIMO) radar with colocated antennas which we call *phased-MIMO radar*. The new technique enjoys the advantages of MIMO radar without sacrificing the main advantage of phased-array radar which is the coherent processing gain at the transmitting side. The essence of the proposed technique is to partition the transmitting array into a number of subarrays that are allowed to overlap. Then, each subarray is used to coherently transmit a waveform which is orthogonal to the waveforms transmitted by other subarrays. Coherent processing gain can be achieved by designing a weight vector for each subarray to form a beam towards a certain direction in space. Moreover, the subarrays are combined jointly to form a MIMO radar resulting in higher resolution capabilities. The substantial improvements offered by the proposed phased-MIMO radar technique as compared to previous techniques are demonstrated analytically and by simulations through analysis of the corresponding beampatterns and achievable output signal-to-noise-plus-interference ratios. Both analytical and simulation results validate the effectiveness of the proposed phased-MIMO radar.

## I. Introduction

Radar technology has been continuously developing over the last 70 years starting from the late 30s of the last century when radar was first invented for defence applications [1]–[3]. The desire for new

\*Supported in parts by the Natural Science and Engineering Research Council (NSERC) of Canada and the Alberta Ingenuity Foundation, Alberta, Canada.

The authors are with the Department of Electrical and Computer Engineering, University of Alberta, 9107-116 St., Edmonton, Alberta, T6G 2V4 Canada. Emails: {hassanie, vorobyov}@ece.ualberta.ca

**Corresponding author:** Sergiy A. Vorobyov, Dept. Elect. and Comp. Eng., University of Alberta, 9107-116 St., Edmonton, Alberta, T6G 2V4, Canada; Phone: +1 780 492 9702, Fax: +1 780 492 1811. Email: vorobyov@ece.ualberta.ca.

more advanced radar technologies has been driven and is dictated by radar's ubiquitous applicability ranging from micro-scale radars applied in biomedical engineering [4], [5] to macro-scale radars used in radioastronomy [6], [7]. To date, previous developments in radar were based on the idea that the signal can be processed coherently at the transmit/receive antenna arrays if the signal coherency is preserved. The corresponding radar technique is well known under the name *phased-array radar* [1], [2].

In the last decade, the development of a new radar paradigm that is best known under the title *multiple-input multiple-output (MIMO) radar* has become the focus of intensive research [8]–[15]. The essence of the MIMO radar concept is to employ multiple antennas for emitting several orthogonal waveforms and multiple antennas for receiving the echoes reflected by the target. The enabling concept for MIMO radar, e.g., the transmission of multiple orthogonal waveforms from different antennas, is usually referred to as the *waveform diversity* [8], [9]. Consequently, the waveform design and optimization has been the main focus of the research in MIMO radar [16]-[20].

Many approaches to MIMO radar have been developed that evolve around the main idea of exploiting the waveform diversity. Based on the array configurations used, MIMO radars can be classified into two main types. The first type uses widely separated transmit/receive antennas to capture the spatial diversity of the target's radar cross section (RCS) (see [9], and references therein). This type assumes an extended target model and, therefore, takes advantages of the properties of the associated spatially-distributed signal model. In this case, the waveform diversity is similar to the diversity concept in wireless communications over fading channels where signals (or their simple modifications) are transmitted over multiple fading links/channels and can be decoded reliably at the receiver due to the fact that it is unlikely that all links/channels undergo unfavorable fading conditions simultaneously [21], [22].

The second MIMO radar type employs arrays of closely spaced transmit/receive antennas to cohere a beam towards a certain direction in space (see [13], and references therein). In this case, the target is usually assumed to be in the far-field and, therefore, the point source signal model is commonly presumed. The waveform diversity, in this case, boils down to increasing the virtual aperture of the receiving array due to the fact that multiple independent waveforms are received by the same receiving array [13].[1]

In this paper, we consider the case of a radar system with colocated antennas. As compared to the phased-array radar, the use of MIMO radar with colocated antennas enables improving angular resolution, increasing the upper limit on the number of detectable targets, improving parameter identifiability, extend-ing the array aperture by virtual sensors, and enhancing the flexibility for transmit/receive beampattern design [13]–[15]. However, the advantages offered by the MIMO radar come at the price of loosing coherent processing gain at the transmitting array offered by the phased-array radar. Hence, the MIMO radar with colocated antennas may suffer from beam-shape loss which leads to performance degradation in the presence of target's RCS fading.

---

[1]It would be, however, more precise to rename the waveform diversity concept as the *multiple independent waveforms* concept in the case of MIMO radar with collocated antennas. We use the traditional terminology hereafter though.

To overcome the aforementioned weakness of the MIMO radar, we develop a new radar technique which combines the advantages of the MIMO radar (such as waveform diversity) with the advantages of the phased-array radar (such as coherent processing).[2] In order to enable such an integration, we partition the transmitting array into a number of subarrays that are allowed to overlap[3]. Then, each subarray is used to coherently transmit a waveform which is orthogonal to the waveforms transmitted by other subarrays. Coherent processing gain can be achieved by designing the weight vector of each subarray to form a beam towards a certain direction in space (the same direction for all subarrays). In parallel, the subarrays are combined jointly to form a MIMO radar resulting in higher resolution capabilities.

The advantages of the new radar technique, that is naturally called *phased-MIMO radar*, over the phased-array and MIMO radars are analyzed in terms of the corresponding beampattern and signal-to-noise-plus-interference ratios (SINRs) expressions. Particularly, the new radar technique:

 (i)   enjoys all the advantages of the MIMO radar, i.e., it enables improving angular resolution, detecting a higher number of targets, improving parameter identifiability, and extending the array aperture;
 (ii)  enables the use of existing beamforming techniques at both the transmitting and the receiving ends;
 (iii) provides the means for designing the overall beampattern of the virtual array;
 (iv)  offers a tradeoff between resolution and robustness against beam-shape loss;
 (v)   offers improved robustness against strong interference.

Our paper is organized as follows. Some background on MIMO radar is presented in Section II, where several concepts needed for developing a new phased-MIMO radar technique are revisited. In Section III, we give a formulation of a new phased-MIMO radar and highlight its advantages and some related design problems. Section IV is devoted to the analysis of the phased-MIMO radar with transmit/receive beamforming. Particularly, we derive the phased-MIMO radar beampattern for the case when conventional transmit/receive beamforming is used, and compare it with the beampatterns of the phased-array and MIMO radars. We also compare the three aforementioned radar techniques in terms of their achievable output SINRs. The possibility of using robust/adaptive beamforming is also discussed. Section V presents our simulation results which show significant performance gains that can be achieved by the phased-MIMO radar technique as compared to the phased-array and MIMO radars. Section VI contains our concluding remarks.

---

[2] An early exposition of this work has been presented in [23].

[3] Partitioning of the transmitting array into subarrays has been also recently used in [24] for signal transmission at different directions from different subarrays. The subarrays, in this case, have to be non-overlapped. Moreover, the problem considered is also totaly different from our problem.

## II. MIMO Radar: Preliminaries

Consider a radar system with a transmitting array equipped with $M$ colocated antennas and a receiving array equipped with $N$ colocated antennas. Both the transmitting and receiving arrays are assumed to be close to each other in space (possibly the same array) so that they see targets at same directions. The $m$th transmitting antenna emits the $m$th element of the waveform vector $\boldsymbol{\phi}(t) \triangleq [\phi_1(t), \ldots, \phi_M(t)]^T$ which satisfies the orthogonality condition

$$\int_{T_0} \boldsymbol{\phi}(t) \boldsymbol{\phi}^H(t) dt = \mathbf{I}_M \tag{1}$$

where $T_0$ is the radar pulse width, $t$ is the time index within the radar pulse, $\mathbf{I}_M$ is the $M \times M$ identity matrix, and $(\cdot)^T$ and $(\cdot)^H$ stand for the transpose and Hermitian transpose, respectively. The total energy transmitted by a MIMO radar system within one radar pulse is given by

$$E_{\text{MIMO}} = \text{trace}\left\{\int_{T_0} \boldsymbol{\phi}(t) \boldsymbol{\phi}^H(t) dt\right\} = M. \tag{2}$$

The $N \times 1$ snapshot vector received by the receiving array can be modeled as

$$\mathbf{x}(t) = \mathbf{x}_s(t) + \mathbf{x}_i(t) + \mathbf{n}(t) \tag{3}$$

where $\mathbf{x}_s(t)$, $\mathbf{x}_i(t)$, and $\mathbf{n}(t)$ are the independent components of the target/source signal, interference/jamming, and sensor noise, respectively. Under point target assumption, the target signal can be written as

$$\mathbf{x}_s(t) = \beta_s \left(\mathbf{a}^T(\theta_s) \boldsymbol{\phi}(t)\right) \mathbf{b}(\theta_s) \tag{4}$$

where $\theta_s$ is the target direction, $\beta_s$ is the complex-valued reflection coefficient of the focal point $\theta_s$, and $\mathbf{a}(\theta)$ and $\mathbf{b}(\theta)$ are the actual transmit and actual receive steering vectors associated with the direction $\theta$.

The returns due to the $m$th transmitted waveform can be recovered by match filtering the received signal to each of the waveforms $\{\phi_m(t)\}_{m=1}^{M}$, i.e.,

$$\mathbf{x}_m \triangleq \int_{T_0} \mathbf{x}(t) \phi_m^*(t) dt, \quad m = 1, \ldots, M \tag{5}$$

where $(\cdot)^*$ denotes the conjugate operator. Then, the $MN \times 1$ virtual data vector can be written as

$$\mathbf{y} \triangleq [\mathbf{x}_1^T \cdots \mathbf{x}_M^T]^T = \beta_s \mathbf{a}(\theta_s) \otimes \mathbf{b}(\theta_s) + \mathbf{y}_{i+n} \quad (6)$$

where $\otimes$ stands for the Kronker product and $\mathbf{y}_{i+n}$ accounts for the interference-plus-noise components. The target signal component in (6) can be expressed as

$$\mathbf{y}_s \triangleq \beta_s \mathbf{v}(\theta_s) \quad (7)$$

where $\mathbf{v}(\theta) \triangleq \mathbf{a}(\theta) \otimes \mathbf{b}(\theta)$ is the $MN \times 1$ steering vector associated with a virtual array of $MN$ sensors[4].

For the special case of uniform linear array (ULA) at the transmitter and the receiver, the $(mN+n)$th entry of the virtual array steering vector $\mathbf{v}(\theta)$ is given by

$$\mathbf{v}_{[mN+n]}(\theta) = e^{-\jmath 2\pi(m d_T \sin\theta + n d_R \sin\theta)}, \quad m=0,\ldots,M-1, \; n=0,\ldots,N-1 \quad (8)$$

where $d_T$ and $d_R$ are the inter-element spacings measured in wavelength for the transmitting and receiving arrays, respectively. It was shown in the literature that the inter-element spacing of the transmitting array can take values higher than a half wavelength without suffering from ambiguity at the receiving end [14]. In particular, if $d_T = N d_R$ is chosen, then the resulting virtual array is a ULA of $MN$ elements spaced $d_R$ wavelength apart from each other and its steering vector simplifies to [14]

$$\mathbf{v}_{[\zeta]}(\theta) = e^{-\jmath 2\pi d_R \zeta \sin\theta}, \quad \zeta \triangleq mN + n = 0,1,\ldots,MN-1. \quad (9)$$

which means that an array with $MN$ effective aperture can be obtained by using $M+N$ antennas. Based on the latter fact, higher resolution and better performance can be achieved by using $\mathbf{y}$ in (6) for detection and estimation purposes. Despite the aforementioned advantages, the MIMO radar has also significant disadvantage as compared to the phased-array radar. Particularly, the above MIMO radar formulation does not allow for coherent processing (beamforming) at the transmitting array and, therefore, it lacks robustness against sensor noise and RCS fading. Moreover, the MIMO radar has $M$ times smaller "clear region" as compared to the phased-array radar [25].

## III. Proposed Phased-MIMO Radar Formulations

In this section, we propose a new formulation for MIMO radar which allows for beamforming at the transmitting and receiving arrays. The new formulation enables a compromise between the coherent processing gain offered by the phased-array radar and the above-mentioned advantages of the MIMO radar, e.g., the waveform diversity. Therefore, we call such new radar technique as phased-MIMO radar.

---

[4]The size extension of the resulting virtual array due to the orthogonality of the transmitted waveforms is traditionally referred to as waveform diversity for MIMO radar with colocated antennas.

The main idea behind our formulations is to partition the transmitting array into $K$ subarrays ($1 \leq K \leq M$) which are allowed to overlap. In general, each subarray can be composed of any number of antennas ranging from 1 to $M$. However, in this paper we adopt the partitioning shown in Fig. 1 where the $k$th subarray is composed of the antennas located at the $k$th up to the $(M-K+k)$th positions, i.e., each subarray consists of $M-K+1$ antennas. We call such partitioning as fully-overlapped, hereafter. It will be shown in Section IV that this way of partitioning has advantages over other ways. All elements of the $k$th subarray are used to coherently emit the signal $\phi_k(t)$ so that a beam is formed towards a certain direction in space, e.g., direction of the target. Then, the beamforming weight vector can be properly designed to maximize the coherent processing gain. At the same time, different waveforms are transmitted by different subarrays.

The signal transmitted by the $k$th subarray can be modeled as

$$\mathbf{s}_k(t) = \sqrt{\frac{M}{K}} \phi_k(t) \mathbf{w}_k^*, \quad k = 1, \ldots, K \tag{10}$$

where $\mathbf{w}_k$ is the unit-norm complex vector of beamforming weights associated with the $k$th subarray. It is worth noting that the energy of $\mathbf{s}_k(t)$ within one radar pulse is given by

$$E_\mathrm{k} = \int_{T_0} \mathbf{s}_k^H(t) \mathbf{s}_k(t) dt = \frac{M}{K} \tag{11}$$

which means that the total transmitted energy for the phased-MIMO radar $E_{\mathrm{PH-MIMO}}$ within one radar pulse is equal to $M$.

The signal reflected by a hypothetical target located at direction $\theta$ in the far-field can be then modeled as

$$r(t, \theta) \triangleq \sqrt{\frac{M}{K}} \beta(\theta) \sum_{k=1}^{K} \mathbf{w}_k^H \mathbf{a}_k(\theta) e^{-\jmath \tau_k(\theta)} \phi_k(t) \tag{12}$$

where $\beta(\theta)$ is the reflection coefficient[5] of the hypothetical target, $\mathbf{a}_k(\theta)$ is the steering vector associated with the $k$th subarray, and $\tau_k(\theta)$ is the time required for the wave to travel across the spatial displacement between the first element of the first subarray and the first element of the $k$th subarray.

Let us introduce the $K \times 1$ transmit coherent processing vector

$$\mathbf{c}(\theta) \triangleq \left[ \mathbf{w}_1^H \mathbf{a}_1(\theta), \ldots, \mathbf{w}_K^H \mathbf{a}_K(\theta) \right]^T \tag{13}$$

---

[5]The reflection coefficient for each target is assumed to be constant during the whole pulse but varies from pulse to pulse.

and the $K \times 1$ waveform diversity vector

$$\mathbf{d}(\theta) \triangleq \left[ e^{-\jmath \tau_1(\theta)}, \ldots, e^{-\jmath \tau_K(\theta)} \right]^T. \tag{14}$$

Then, the reflected signal (12) can be rewritten as

$$r(t, \theta) = \sqrt{\frac{M}{K}} \beta(\theta) \left( \mathbf{c}(\theta) \odot \mathbf{d}(\theta) \right)^T \boldsymbol{\phi}_K(t) \tag{15}$$

where where $\boldsymbol{\phi}_K(t) = [\phi_1(t), \ldots, \phi_K(t)]$ is the $K \times 1$ vector of waveforms and $\odot$ stands for the Hadamard (element-wise) product.

Assuming that the target of interest is observed in the background of $D$ interfering targets with reflection coefficients $\{\beta_i\}_{i=1}^{D}$ and locations $\{\theta_i\}_{i=1}^{D}$, the $N \times 1$ received complex vector of array observations can be written as

$$\mathbf{x}(t) = r(t, \theta_s) \mathbf{b}(\theta_s) + \sum_{i=1}^{D} r(t, \theta_i) \mathbf{b}(\theta_i) + \mathbf{n}(t) \tag{16}$$

where, $r(t, \theta_i)$ is defined as in (15). By mach-filtering $\mathbf{x}(t)$ to each of the waveforms $\{\phi_k\}_{k=1}^{K}$ (e.g., as in (5)), we can form the $KN \times 1$ virtual data vector

$$\begin{aligned} \mathbf{y} &\triangleq [\mathbf{x}_1^T \cdots \mathbf{x}_K^T]^T \\ &= \sqrt{\frac{M}{K}} \beta_s \mathbf{u}(\theta_s) + \sum_{i=1}^{D} \sqrt{\frac{M}{K}} \beta_i \mathbf{u}(\theta_i) + \tilde{\mathbf{n}} \end{aligned} \tag{17}$$

where the $KN \times 1$ vector

$$\mathbf{u}(\theta) \triangleq (\mathbf{c}(\theta) \odot \mathbf{d}(\theta)) \otimes \mathbf{b}(\theta) \tag{18}$$

is the virtual steering vector associated with direction $\theta$ and $\tilde{\mathbf{n}}$ is the $KN \times 1$ noise term whose covariance is given by $\tilde{\mathbf{R}}_{\mathbf{n}} = \sigma_n^2 \mathbf{I}_{KN}$.

It is worth noting that if $K = 1$ is chosen, i.e. if the whole transmitting array is considered as one subarray and only one waveform is emitted, then the signal model (17) simplifies to the signal model for the conventional phased-array radar [26]

$$\mathbf{y} = \sqrt{M} \beta_s \mathbf{u}(\theta_s) + \sum_{i=1}^{D} \sqrt{M} \beta_i \mathbf{u}(\theta_i) + \tilde{\mathbf{n}} \tag{19}$$

while the virtual steering vector given in (18) simplifies to

$$\mathbf{u}(\theta) = \left[ \mathbf{w}^H \mathbf{a}(\theta) \right] \cdot \mathbf{b}(\theta) \tag{20}$$

where $\mathbf{w}^H \mathbf{a}(\theta)$ is the uplink coherent processing gain of the conventional phased-array radar towards the direction $\theta$ and $\mathbf{w}$ is the $M \times 1$ uplink beamformer weight vector. In this case, the received data vector $\mathbf{y}$ is of dimension $N \times 1$ which explains the low resolution performance of the phased-array radar.

On the other hand, if $K = M$ is chosen, then the signal model (17) simplifies to (6) which is the signal model for the MIMO radar without array partitioning. In this case, the $MN \times 1$ data vector $\mathbf{y}$ enables the highest possible resolution at the price of having no coherent processing gain at the transmitting side.

The proposed MIMO radar formulation (17) combines the benefits/advantages of the phased-array radar (19)–(20) and the MIMO radar (6). In particular, the new formulation enjoys many interesting characteristics. First, it enjoys all advantages of MIMO radar, i.e., it enables improving angular resolution, detecting a higher number of targets, improving parameter identifiability, extending the array aperture by virtual sensors. Second, it allows for uplink beamforming at the transmitting side and, therefore, it enables optimization/maximization of the coherent processing gain and controlling/minimizing the total transmit power. Third, it provides the means for designing the overall beampattern of the virtual array. In other words, it empowers optimization of the overall beampattern of the virtual array by designing the transmit/receive beamforming weights jointly. Fourth, the proposed formulation offers a tradeoff between resolution and robustness against beam-shape loss by properly selecting the number of subarrays used. Finally, it offers a tradeoff between improvements in performance and the required computational complexity.

By inspecting (10), it is interesting to observe that different antennas transmit linear combinations of the base orthogonal waveforms $\{\phi_k(t)\}_{k=1}^K$. Let $\{\psi_m(t)\}_{m=1}^M$ be the signals transmitted by the $M$ transmit antennas. These signals need not be orthogonal and can be expressed as

$$\boldsymbol{\psi(t)} \triangleq \sqrt{\frac{M}{K}} \mathbf{W}^* \boldsymbol{\phi}_K(t) \tag{21}$$

where $\boldsymbol{\psi(t)} \triangleq [\psi_1(t), \ldots, \psi_M(t)]^T$ is the $M \times 1$ vector of transmitted signals, $\mathbf{W} \triangleq [\tilde{\mathbf{w}}_1 \ldots, \tilde{\mathbf{w}}_K]$ is the weight matrix of dimension $M \times K$, and

$$\tilde{\mathbf{w}}_k \triangleq [\mathbf{0}_{[k-1]}^T, \mathbf{w}_k^T, \mathbf{0}_{[K-k]}^T]^T \tag{22}$$

is the $M \times 1$ weight vector with $\mathbf{0}_{[k-1]}^T$ denoting the vector of $k - 1$ zeros. Therefore, the proposed phased-MIMO radar can also be interpreted in terms of the MIMO radar with non-orthogonal antenna-wise waveforms, but orthogonal subarray-wise waveforms. Note that because of the inherent structure, the optimal processing of subarray-wise orthogonal waveforms is maintained at the receiving array, where the waveforms orthogonality is required for optimal detection [17].

The power of the target at direction $\theta$ can be expressed as

$$P(\theta) = \mathrm{E}\left\{\left|\sqrt{\frac{M}{K}}\beta(\theta)\mathbf{a}^T(\theta)\boldsymbol{\psi}\right|^2\right\}$$
$$= \frac{M}{K}\sigma^2(\theta)\left\|\mathbf{W}^H\mathbf{a}(\theta)\right\|^2 \tag{23}$$

where $\mathrm{E}\{\cdot\}$ denotes the expectation operator and $\sigma^2(\theta) \triangleq \mathrm{E}\{|\beta(\theta)|^2\}$. Note that the representation (21)–(23) is of special importance when it is required to constrain the transmitted power per each individual antenna (i.e., power of the signals $\{\psi_m\}_{m=1}^M$). However, due to space limitations in this paper, we only impose constrains on the total transmitted power. In other words, we always assume that the total transmitted energy within one radar pulse equals to $M$. Moreover, in our formulation of the phased-MIMO radar, the subarray selection is not adaptive, i.e., it is independent on the reflection coefficient $\beta(\theta)$. Therefore, extensions of the proposed technique are possible.

## IV. PHASED-MIMO RADAR TRANSMIT/RECEIVE BEAMFORMING

At the transmitting array, existing uplink beamforming techniques lend themselves easily to design the weight vectors $\{\mathbf{w}_k\}_{k=1}^K$ for different subarrays such that certain beampattern and/or transmit power requirements are satisfied. In this section, we apply and analyze the transmit/receive beamforming techniques for the proposed phased-MIMO radar (17). We also compare the phased-MIMO radar to the phased-array radar (19) and the MIMO radar (6) in terms of their transmit-receive beampatterns and achievable SINRs. We consider in details the case when non-adaptive transmit/receive beamforming techniques are used. Moreover, we briefly discuss the possibility of applying adaptive transmit/receive beamforming for the proposed phased-MIMO radar.

### A. Non-adaptive Transmit/Receive Beamforming

The signal-to-noise ratio (SNR) gain of the phased-array radar is proportional to the quantity $|\mathbf{w}^H\mathbf{a}(\theta_s)|$. Using the Cauchy-Schwarz inequality, we have $|\mathbf{w}^H\mathbf{a}(\theta_s)| \leq \|\mathbf{w}\| \cdot \|\mathbf{a}(\theta_s)\|$, where equality holds when $\mathbf{w} = \mathbf{a}(\theta_s)$ which is referred to as the conventional beamformer. In the case when a single source signal is observed in the background of white Gaussian noise, the conventional beamformer is known to be optimal in the sense that it provides the highest possible output SNR gain. For this reason and due to its simplicity, the conventional (non-adaptive) beamformer has been widely used in antenna array systems [27]. Therefore, we use the conventional beamforming at both the transmitting and receiving arrays of the phased-MIMO radar system and derive expressions for transmit/receive beampattern and output SINR. We

also analyze and compare the phased-MIMO radar beampattern and SINR expressions to the analogous expressions for the phased-array and MIMO radars.

Without loss of generality, we can assume that all subarrays have equal aperture, i.e. have the same number of elements. For conventional uplink beamforming, the beamformer wight vectors are given by

$$\mathbf{w}_k = \frac{\mathbf{a}_k(\theta_s)}{\|\mathbf{a}_k(\theta_s)\|}, \quad k = 1, \ldots, K. \tag{24}$$

At the receiving array, the conventional beamformer is applied to the virtual array and, therefore, the $KN \times 1$ receive beamformer weight vector is given by

$$\mathbf{w}_d \triangleq \mathbf{u}(\theta_s) = [\mathbf{c}(\theta_s) \odot \mathbf{d}(\theta_s)] \otimes \mathbf{b}(\theta_s). \tag{25}$$

Let $G(\theta)$ be the normalized phased-MIMO radar beampattern, that is

$$G(\theta) \triangleq \frac{\left|\mathbf{w}_d^H \mathbf{u}(\theta)\right|^2}{\left|\mathbf{w}_d^H \mathbf{u}(\theta_s)\right|^2} = \frac{\left|\mathbf{u}^H(\theta_s)\mathbf{u}(\theta)\right|^2}{\|\mathbf{u}(\theta_s)\|^4} \tag{26}$$

To simplify the derivation and analysis of the beampattern and SINR expressions for the phased-MIMO radar, we assume a ULA, i.e.,

$$\mathbf{a}_1^H(\theta_s)\mathbf{a}_1(\theta) = \ldots = \mathbf{a}_K^H(\theta_s)\mathbf{a}_K(\theta). \tag{27}$$

Hence, the beampattern (26) can be reformulated as follows

$$G_K(\theta) = \frac{\left|\mathbf{a}_K^H(\theta_s)\mathbf{a}_K(\theta)\left[(\mathbf{d}(\theta_s) \otimes \mathbf{b}(\theta_s))^H(\mathbf{d}(\theta) \otimes \mathbf{b}(\theta))\right]\right|^2}{\|\mathbf{a}_K^H(\theta_s)\|^4 \|\mathbf{d}(\theta_s) \otimes \mathbf{b}(\theta_s)\|^4} \tag{28}$$

$$= \frac{\left|\mathbf{a}_K^H(\theta_s)\mathbf{a}_K(\theta)\right|^2 \left|\mathbf{d}^H(\theta_s)\mathbf{d}(\theta)\right|^2 \left|\mathbf{b}^H(\theta_s)\mathbf{b}(\theta)\right|^2}{\left\|\mathbf{a}_K^H(\theta_s)\right\|^4 \|\mathbf{d}(\theta_s)\|^4 \|\mathbf{b}(\theta_s)\|^4} \tag{29}$$

where (29) is obtained from (28) using the fact that

$$(\mathbf{d}(\theta_s) \otimes \mathbf{b}(\theta_s))^H(\mathbf{d}(\theta) \otimes \mathbf{b}(\theta)) = \mathbf{d}^H(\theta_s)\mathbf{d}(\theta)\mathbf{b}^H(\theta_s)\mathbf{b}(\theta). \tag{30}$$

Let us consider three different cases for partitioning he transmitting array.

Case 1: Each subarray is composed of all $M$ transmit antennas, i.e., each subarray is the same as the whole array. In this case, we have $\mathbf{a}_K(\theta) = \mathbf{a}(\theta)$ and $\mathbf{d}(\theta) = \mathbf{1}$, where $\mathbf{1}$ is the vector of ones. In this case, the beampattern (29) boils down to

$$G_K(\theta) = \frac{\left|\mathbf{a}^H(\theta_s)\mathbf{a}(\theta)\right|^2 \left|\mathbf{b}^H(\theta_s)\mathbf{b}(\theta)\right|^2}{M^2 N^2} \tag{31}$$

which is the beampattern of a phased-array of $M$ transmit and $N$ receive antennas.

Case 2: The $K$ subarrays are non-overlapped and, therefore, each subarray consists of $M/K$ antennas. This type of partitioning is adopted in [24]. In this case, the $k$th element of vector $\mathbf{d}(\theta)$ corresponds to the $(kM/K+1)$th element of $\mathbf{a}(\theta)$ and the $M/K \times 1$ vectors $\mathbf{a}_1(\theta) = \ldots = \mathbf{a}_K(\theta)$ contain the first $M/K$ elements of $\mathbf{a}(\theta)$ (see (27)). Hence, it is easy to show that $\left|\mathbf{a}_K^H(\theta_s)\mathbf{a}_K(\theta)\right|^2 \left|\mathbf{d}^H(\theta_s)\mathbf{d}(\theta)\right|^2 = \left|\mathbf{a}^H(\theta_s)\mathbf{a}(\theta)\right|^2$. Therefore, the beampattern (29) boils down again to

$$G_K(\theta) = \frac{\left|\mathbf{a}^H(\theta_s)\mathbf{a}(\theta)\right|^2 \left|\mathbf{d}^H(\theta_s)\mathbf{d}(\theta)\right|^2 \left|\mathbf{b}^H(\theta_s)\mathbf{b}(\theta)\right|^2}{(M/K)^2 K^2 N^2}$$
$$= \frac{\left|\mathbf{a}^H(\theta_s)\mathbf{a}(\theta)\right|^2 \left|\mathbf{b}^H(\theta_s)\mathbf{b}(\theta)\right|^2}{M^2 N^2}. \quad (32)$$

Case 3: In this case, we consider our method of partitioning shown in Fig. 1. Recall that each subarray consists of $M-K+1$ transmit antennas. Due to full-overlap, it is easy to show that $\mathbf{d}(\theta) = [\mathbf{a}_{[1]}(\theta), \ldots, \mathbf{a}_{[K]}(\theta)]^T$, where $\mathbf{a}_{[k]}(\theta)$ is the $k$th element of the transmit steering vector $\mathbf{a}(\theta)$. Therefore, the phased-MIMO radar beampattern (29) can be rewritten as

$$G_K(\theta) = \frac{\left|\mathbf{a}_K^H(\theta_s)\mathbf{a}_K(\theta)\right|^2}{(M-K+1)^2} \cdot \frac{\left|\mathbf{d}^H(\theta_s)\mathbf{d}(\theta)\right|^2}{K^2} \cdot \frac{\left|\mathbf{b}^H(\theta_s)\mathbf{b}(\theta)\right|^2}{N^2} \quad (33)$$

where the facts that $\|\mathbf{a}_K(\theta_s)\|^2 = M-K+1$, $\|\mathbf{d}(\theta_s)\|^2 = K$, and $\|\mathbf{b}(\theta_s)\|^2 = N$ are used.

Let $C_K(\theta) \triangleq \frac{|\mathbf{a}_K^H(\theta_s)\mathbf{a}_K(\theta)|^2}{(M-K+1)^2}$, $D_K(\theta) \triangleq \frac{|\mathbf{d}^H(\theta_s)\mathbf{d}(\theta)|^2}{K^2}$, and $R(\theta) \triangleq \frac{|\mathbf{b}^H(\theta_s)\mathbf{b}(\theta)|^2}{N^2}$ be the transmit (uplink) beampattern, the waveform diversity beampattern, and the receive (downlink) beampattern, respectively. Then, the phased-MIMO radar beampattern can be seen as the product of three individual beampatterns, i.e.,

$$G_K(\theta) = C_K(\theta) \cdot D_K(\theta) \cdot R(\theta). \quad (34)$$

By inspecting (34), we draw the following observations.

- The first two terms $C_K(\theta)$ and $D_K(\theta)$ of the product (34) are dependent on the number of subarrays $K$ while the third term $R(\theta)$ is independent on $K$. Hence, the beampattern analysis for the phased-MIMO radar will focus on the first two terms.

- The beampattern expression for the phased-array radar can be deduced form (34) by substituting $K=1$. Hence, we obtain

$$G_{\text{PH}}(\theta) = C_1(\theta) \cdot D_1(\theta) \cdot R(\theta) = C_1(\theta) \cdot R(\theta) \quad (35)$$

where $C_1(\theta) = \frac{|\mathbf{a}^H(\theta_s)\mathbf{a}(\theta)|^2}{M^2}$ and $D_1(\theta) = 1$. Note that this case is different from Case 1 for array partitioning mentioned above because only one waveform is transmitted in this case while a mixture of any $K$ orthogonal waveforms is transmitted in Case 1.

- The MIMO radar beampattern expression can be obtained form (34) by substituting $K = M$ which results in

$$G_{\text{MIMO}}(\theta) = C_M(\theta) \cdot D_M(\theta) \cdot R(\theta) = D_M(\theta) \cdot R(\theta) \tag{36}$$

where $C_M(\theta) = 1$ and $D_M(\theta) = \frac{|\mathbf{a}^H(\theta_s)\mathbf{a}(\theta)|^2}{M^2}$.

Comparing (34), (35), and (36) to each other, we notice that the phased-array and the MIMO radar have the same overall beampattern, i.e.,

$$G_{\text{PH}}(\theta) = G_{\text{MIMO}}(\theta) = \frac{|\mathbf{a}^H(\theta_s)\mathbf{a}(\theta)|^2}{M^2} R(\theta). \tag{37}$$

However, they have different uplink and waveform diversity beampatterns and, therefore, different gains. The phased-array radar has the highest possible transmit coherent processing gain at the price of no diversity gain, while the MIMO radar has the highest waveform diversity gain at the price of no transmit coherent processing gain. On the other hand, the beampattern of the phased-MIMO radar enjoys two interesting properties which are given in the following two propositions.

***Proposition 1*:** If the phased-MIMO radar is formed by partitioning a ULA into $K$ fully-overlapped subarrays, then the transmit-receive beampattern equals to the transmit-receive beampattern of the phased-MIMO radar formed by partitioning the same ULA into $M-K+1$ subarrays, i.e.,

$$|G_K(\theta)| = |G_{M-K+1}(\theta)|. \tag{38}$$

*Proof:* Noting that $\mathbf{a}_K(\theta)$ is of dimension $(M-K+1) \times 1$ and $\mathbf{d}(\theta)$ is of dimension $K \times 1$, the proof is readily obtained by substituting $K = M-K+1$ in (33) and exchanging $\mathbf{a}_K(\theta)$ for $\mathbf{d}(\theta)$ and $\mathbf{a}_K(\theta_s)$ for $\mathbf{d}(\theta_s)$. ∎

***Proposition 2*:** The transmit-receive beampattern of the phased-MIMO radar with $K$ subarrays has lower highest sidelobe level than the transmit-receive beampattern of the phased-array radar, i.e.,

$$\max_{\theta \in \bar{\Theta}} C_K(\theta) \cdot D_K(\theta) \cdot R(\theta) \leq \max_{\theta \in \bar{\Theta}} C_1(\theta) \cdot R(\theta) \tag{39}$$

where $\bar{\Theta}$ is a continuum of all spatial angels within the sidelobe area.

*Proof:* See Appendix A.

Based on Proposition 2, the phased-MIMO radar enjoys better robustness against interfering targets located in the sidelobe area as compared to the phased-array and MIMO radars. This also implies that the proposed partitioning scheme is superior to other types of array partitioning. To illustrate this, let us examine the optimal output SINR for all three radar techniques. The output SINR of the phased-MIMO radar can be defined as

$$\text{SINR}_{\text{PH-MIMO}} \triangleq \frac{\frac{M}{K}\sigma_s^2 |\mathbf{w}_d^H \mathbf{u}(\theta_s)|^2}{\mathbf{w}_d^H \mathbf{R}_{i+n} \mathbf{w}_d} \tag{40}$$

where the interference-plus-noise covariance matrix is given by

$$\begin{aligned} \mathbf{R}_{i+n} &= \mathrm{E}\left\{\mathbf{y}_{i+n}\mathbf{y}_{i+n}^H\right\} \\ &= \sum_{i=1}^{D} \frac{M}{K}\sigma_i^2 \mathbf{u}(\theta_i)\mathbf{u}^H(\theta_i) + \sigma_n^2 \mathbf{I}. \end{aligned} \tag{41}$$

Substituting $\mathbf{w}_d = \mathbf{u}(\theta_s)$ in (40), the output SINR for the phased-MIMO radar can be rewritten as

$$\begin{aligned} \text{SINR}_{\text{PH-MIMO}} &= \frac{\frac{M}{K}\sigma_s^2 \left|\mathbf{u}^H(\theta_s)\mathbf{u}(\theta_s)\right|^2}{\mathbf{u}^H(\theta_s)\mathbf{R}_{i+n}\mathbf{u}(\theta_s)} \\ &= \frac{\frac{M}{K}\sigma_s^2 \left|\|\mathbf{a}_K(\theta_s)\|^2 \|\mathbf{d}(\theta_s) \otimes \mathbf{b}(\theta_s)\|^2\right|^2}{\mathbf{u}^H(\theta_s)\mathbf{R}_{i+n}\mathbf{u}(\theta_s)} \\ &= \frac{\frac{M}{K}\sigma_s^2 \left|\|\mathbf{a}_K(\theta_s)\|^2 \|\mathbf{d}(\theta_s)\|^2 \|\mathbf{b}(\theta_s)\|^2\right|^2}{\mathbf{u}^H(\theta_s) \left(\sum_{i=1}^{D} \frac{M}{K}\sigma_i^2 \mathbf{u}(\theta_i)\mathbf{u}^H(\theta_i) + \sigma_n^2 \mathbf{I}\right) \mathbf{u}(\theta_s)} \\ &= \frac{\frac{M}{K}\sigma_s^2 (M - K + 1)^2 K^2 N^2}{\sum_{i=1}^{D} \frac{M}{K}\sigma_i^2 \left|\mathbf{u}^H(\theta_s)\mathbf{u}(\theta_i)\right|^2 + \sigma_n^2 (M - K + 1)KN}. \end{aligned} \tag{42}$$

Then, substituting $K = 1$ in (42), we obtain the output SINR for the phased-array radar as

$$\text{SINR}_{\text{PH}} = \frac{\sigma_s^2 M^2 N^2}{\sum_{i=1}^{D} \sigma_i^2 \left|\mathbf{a}^H(\theta_s)\mathbf{a}(\theta_i)\right|^2 \left|\mathbf{b}(\theta_s)^H \mathbf{b}(\theta_i)\right|^2 + \sigma_n^2 N}. \tag{43}$$

Similarly, the output SINR for the MIMO radar can be obtained by substituting $K = M$ in (42) as

$$\text{SINR}_{\text{MIMO}} = \frac{\sigma_s^2 M^2 N^2}{\sum_{i=1}^{D} \sigma_i^2 \left|\mathbf{a}^H(\theta_s)\mathbf{a}(\theta_i)\right|^2 \left|\mathbf{b}(\theta_s)^H \mathbf{b}(\theta_i)\right|^2 + \sigma_n^2 MN}. \tag{44}$$

For the sake of comparison, we analyze the output SINRs given in (42)–(44) for the following two cases.

*1) Dominant noise power:* If the target is observed in the background of few weak interferers which are well separated from the target, then the interference-to-noise power can be attributed to the noise term only. In such a case, the SINR for the phased-array radar simplifies to

$$\text{SINR}_{\text{PH}} \simeq \frac{\sigma_s^2 M^2 N}{\sigma_n^2} \tag{45}$$

while the SINR for the MIMO radar simplifies to

$$\text{SINR}_{\text{MIMO}} \simeq \frac{\sigma_s^2 MN}{\sigma_n^2}. \tag{46}$$

Comparing (45) and (46), we observe that

$$\text{SINR}_{\text{PH}} = M \cdot \text{SINR}_{\text{MIMO}} \qquad (47)$$

which means that the phased-array radar is more robust to background noise as compared to the MIMO radar. On the other hand, the SINR expression for the phased-MIMO radar boils down to

$$\begin{aligned}
\text{SINR}_{\text{PH}-\text{MIMO}} &\simeq \frac{\sigma_s^2 M(M-K+1)^2 KN^2}{\sigma_n^2(M-K+1)KN} \\
&= \frac{M-K+1}{M} \cdot \frac{\sigma_s^2 M^2 N}{\sigma_n^2} \\
&= \eta \cdot \text{SINR}_{\text{PH}} \qquad (48)
\end{aligned}$$

where $\eta \triangleq (M-K+1)/M$ is the ratio of the phased-MIMO radar SINR to the phased-array radar SINR. It is worth noting that $1/M \leq \eta \leq 1$. The dependance of $\eta$ on $K$ shows that the SNR gain of the phased-MIMO radar linearly decreases by increasing $K$. At the same time, larger $K$ provides larger dimension of the extended virtual array. This depicts the tradeoff between SNR gain and high-resolution capabilities.

*2) Dominant interference:* If the target is observed in the background of strong interference, then we can fairly consider the noise power to be negligible as compared to the interference power. By neglecting the noise term in (43) and (44), we obtain that

$$\begin{aligned}
\text{SINR}_{\text{PH}} &\simeq \frac{\sigma_s^2(\theta_s) M^2 N^2}{\sigma_i^2 |\mathbf{a}^H(\theta_s)\mathbf{a}(\theta_i)|^2 |\mathbf{b}^H(\theta_s)\mathbf{b}(\theta_i)|^2} \\
&\simeq \text{SINR}_{\text{MIMO}} \qquad (49)
\end{aligned}$$

which means that both the phased-array and MIMO radars have the same robustness against interference. On the other hand, the SINR of the phased-MIMO radar can be analyzed by reformulating (42) as follows

$$\text{SINR}_{\text{PH}-\text{MIMO}} = \frac{\frac{M}{K}\sigma_s^2(M-K+1)^2 K^2 N^2}{\sum_{i=1}^{D} \frac{M}{K}\sigma_i^2 |\mathbf{u}^H(\theta_s)\mathbf{u}(\theta_i)|^2 + \sigma_n^2(M-K+1)KN} \qquad (50)$$

$$= \frac{\sigma_s^2 M^2 N^2}{\frac{M^2}{(M-K+1)^2 K^2}\sum_{i=1}^{D} \sigma_i^2 |\mathbf{u}^H(\theta_s)\mathbf{u}(\theta_i)|^2 + \sigma_n^2 \frac{MN}{(M-K+1)}} \qquad (51)$$

$$\simeq \frac{\sigma_s^2 M^2 N^2}{\frac{M^2}{(M-K+1)^2 K^2}\sum_{i=1}^{D} \sigma_i^2 |\mathbf{u}^H(\theta_s)\mathbf{u}(\theta_i)|^2}. \qquad (52)$$

where (51) is obtained by multiplying the numerator and denominator of (50) by $M/\bigl(K(M-K+1)^2\bigr)$.

Using (27) and the equality (30), the SINR expression (52) can be rewritten as

$$\text{SINR}_{\text{PH-MIMO}} \simeq \frac{\sigma_s^2 M^2 N^2}{\frac{M^2}{(M-K+1)^2 K^2} \sum_{i=1}^{D} \sigma_i^2 \left|\mathbf{u}^H(\theta_s)\mathbf{u}(\theta_i)\right|^2}$$

$$= \frac{\sigma_s^2(\theta_s) M^2 N^2}{\frac{M^2}{(M-K+1)^2 K^2} \sum_{i=1}^{D} \sigma_i^2 \left|\mathbf{a}_K^H(\theta_s)\mathbf{a}_K(\theta_i)\right|^2 \left|\mathbf{d}^H(\theta_s)\mathbf{d}(\theta_i)\right|^2 \left|\mathbf{b}^H(\theta_s)\mathbf{b}(\theta_i)\right|^2}. \quad (53)$$

It is worth noting that $\frac{M}{(M-K+1)K} \leq 1$ because $(M-K+1)K - M = (M-K)(K-1) \geq 0$. Using this fact and the fact that the phased-MIMO radar has lower highest sidlobe level than the phased-array radar (see Proposition 2), we conclude that

$$\text{SINR}_{\text{PH-MIMO}} \geq \text{SINR}_{\text{PH}} \quad (54)$$

which means that the phased-MIMO radar is capable of providing better SINR performance as compared to the phased-array and MIMO radars. This observation will be verified later by using simulation examples as well.

*B. Robust/Adaptive Beamforming*

To control/minimize the transmitted power, we can resort to robust uplink beamforming. One meaningful approach is to minimize the norm of the beamformer weight vector while upper-bounding the sidelobe levels. Mathematically, this robust uplink beamformer can be formulated as follows

$$\begin{aligned} \min_{\mathbf{w}_k} \quad & \|\mathbf{w}_k\|^2 \\ \text{s. t.} \quad & \mathbf{w}_k^H \mathbf{a}_k(\theta_s) = \mathbf{a}_{[k]}(\theta_s) \\ & \|\mathbf{w}_k^H \mathbf{a}_k(\theta)\| \leq \delta \quad \forall \theta \in \bar{\Theta} \end{aligned} \quad (55)$$

where $\delta$ is the parameter of user choice used to upper-bound the sidelobe levels.

It is worth noting that transmit coherent processing gain offered by either (24) or (55) will be smaller than the transmit coherent processing gain of the phased-array radar. This is due to the fact that the effective array aperture of each transmitting subarray is smaller than the effective array aperture of the whole array. This also makes the main beam associated with each subarray wider than the main beam associated with the phased-array radar. However, this natural drop in performance of transmit beamforming is the price paid for the many benefits that are gained at the receiving end.

It is also possible to use adaptive processing techniques which aim at maximizing the output SINR at the receiving array. Hence, we resort to the famous minimum variance distortionless response (MVDR) beamformer [27]. The essence of the MVDR beamformer is to minimize the interference-plus-noise power while maintaining a distortionless response towards the direction of the target of interest. This can be expressed as the following optimization problem

$$\min_{\mathbf{w}_R} \ \mathbf{w}_R^H \mathbf{R}_{i+n} \mathbf{w}_R \quad \text{subject to} \quad \mathbf{w}_R^H \mathbf{u}(\theta_s) = 1 \tag{56}$$

where $\mathbf{w}_R$ is the $KN \times 1$ receive beamforming weight vector. The solution to (56) is given by [27]

$$\mathbf{w}_R = \frac{\mathbf{R}_{i+n}^{-1} \mathbf{u}(\theta_s)}{\mathbf{u}^H(\theta_s) \mathbf{R}_{i+n}^{-1} \mathbf{u}(\theta_s)}. \tag{57}$$

In practice, the matrix $\mathbf{R}_{i+n}$ is unavailable and, therefore, the sample covariance matrix $\hat{\mathbf{R}} \triangleq \sum_{n=1}^{N} \mathbf{y}_n \mathbf{y}_n^H$ is used, where $\{\mathbf{y}_n\}_{n=1}^{N}$ are data snapshots which can be collected from $N$ different radar pulses within a coherent processing interval. It is worth noting that the target signal component is present in $\hat{\mathbf{R}}$. An alternative way to obtain a target signal-free sample covariance matrix is to collect the data snapshots $\{\mathbf{y}_n\}_{n=1}^{N}$ for $N$ different range bins [14], [28]. The latter is used in our simulations. Also note that the avenue for using robust adaptive beamforming techniques [29]-[31] is also opened.

## V. SIMULATION RESULTS

In our simulations, we assume a ULA of $M=10$ omnidirectional antennas used for transmitting the baseband waveforms $\left\{ s_k(t) = e^{j2\pi \frac{k}{T_0} t} \right\}_{k=1}^{K}$. We also assume a ULA of $N=10$ omnidirectional antennas spaced half a wave length apart from each other at the receiving end. The additive noise is modeled as a complex Gaussian zero-mean spatially and temporally white random sequence that has identical variances in each array sensor. We assume two interfering targets located at directions $-30°$ and $-10°$ except in Example 4 where one spatially distributed interference is assumed. The target of interest is assumed to reflect a plane-wave that impinges on the array from direction $\theta_s = 10°$. In all our simulation examples we compare the proposed phased-MIMO radar (17) with the phased-array radar (19) and the MIMO radar (6). For the phased-MIMO radar, we always used $K = 5$ subarrays which are assumed to be fully overlapped. In some examples, we compare different radar techniques to each other in terms of their transmit/receive beampatterns, while in other examples, the performance of the aforementioned radar techniques is compared in terms of the output SINRs. The sample covariance matrix is computed based on $N = 100$ data snapshots (i.e., 100 range bins) for all methods tested. Note that for the MIMO radar technique, the sample covariance matrix is of size $100 \times 100$. To avoid the effect of low sample size, the diagonal loading (DL) of $10\mathbf{I}$ is used when solving (57). Note that this DL is used not only for the MIMO radar but also for the other two radar techniques tested for the reason of fair comparison. In all examples, output SINRs are computed based on 100 independent simulation runs for all methods tested.

*A. Non-Adaptive Transmit/Receive Beamforming*

*Example 1: Non-adaptive transmit/receive beampattern without spatial transmit aliasing*

In the first example, we examine the transmit/receive beampattern of the transmit/receive beamformer (24)–(25), for the case when the transmitting antennas are located half a wavelength apart, i.e. $d_T = 0.5$ wavelength. Figs. 2 and 3 show the transmit beampatterns and the waveform diversity beampatterns, respectively, for all three radar techniques tested, while Fig. 4 shows the overall transmit/receive beampatterns for the same techniques.

From Fig. 2, we can see that the phased-array radar has the typical conventional beampattern with mainlobe (of width $\pi/M$) centered at $\theta_s$ while the MIMO radar has flat (0 dB) transmitting gain. On the other hand, the phased-MIMO transmit beampattern is characterized by the aperture (actual size) of the individual subarrays. Since the aperture of the subarrays is always smaller than the aperture of the whole array, the transmit beampattern of the phased-MIMO radar represents a tradeoff between the beampatterns of the MIMO and phased-array radars. As we can see in Fig. 2, the reduction in the subarray aperture results in the beampattern of the phased-MIMO radar with a wider main beam and a little higher sidelobe levels as compared to the beampattern of the phased-array radar. This small loss in beampattern shape is repaid at a greater gain in the waveform diversity beampattern as shown in Fig. 3. It is noted from Fig. 3 that the phased-array radar has no waveform diversity gain (0 dB flat pattern), while the waveform diversity beampatterns of the MIMO and phased-MIMO radars are equivalent to conventional beampatterns offered by an $M$ and $K$ elements virtual arrays, respectively. Because $K \leq M$, the waveform diversity beampattern of the phased-MIMO radar has a wider mainlobe and higher sidelobe levels as compared to the waveform diversity beampattern of the MIMO radar. However, one can see in Fig. 4 that the overall transmit/receive beampattern shape for the proposed phased-MIMO radar is significantly improved as compared to the beampatterns of the phased-array and MIMO radars. Particularly, it is worth noting that the overall beampattern of the proposed phased-MIMO radar is proportional to the multiplication of the transmit and the waveform diversity beampatterns (i.e., proportional to the summation of the curves in Figs. 2 and 3 in dB). We also can observe from Fig. 4 that the phased-array and MIMO radars have exactly the same overall transmit/receive beampatterns. At the same time, the phased-MIMO radar has lower sidelobe levels as compared to both the phased-array and MIMO radars.

*Example 2: Non-adaptive transmit/receive beampattern with spatial transmit aliasing*

In this example, we investigate the non-adaptive transmit/receive beampattern for the case when the transmitting antennas are located more than a half wavelength apart from each other. In particular, the case $d_T = 5d_R$ ($d_R = 0.5$ wavelength) is chosen and the corresponding non-adaptive beamforming based beampatterns for the phased-array, MIMO, and phased-MIMO radars are plotted.

The transmit beampattern is shown in Fig. 5 for all methods tested. All curves have similar trends to their counterparts in Fig. 2 except that, due to spatial aliasing in the transmit mode, each beampattern is repeated 5 times within the spatial domain $[-\pi/2, \pi/2]$. The reason is that the interelement spacing used is 5 times half a wavelength, i.e., 5 times the critical spatial sampling spacing. Similarly, the diversity beampattern exhibits spatial aliasing for all methods as shown in Fig. 6. As in the previous example, we observe from Figs. 5 and 6 that the phase-array radar exhibits high transmit coherent gain at the price of 0 dB waveform diversity gain while the MIMO radar exhibits the opposite, i.e., high waveform diversity gain at the price of flat 0 dB transmit coherent processing gain. At the same time, the phased-MIMO radar is shown to have non-flat both coherent processing and waveform diversity gains. The overall transmit/receive beampattern is shown in Fig. 7. It can be seen from this figure that the phased-array and MIMO radars have exactly the same receive beampatterns, while the proposed phased-MIMO radar enjoys much lower sidelobe levels as compared to the other radar techniques. Hence, the proposed phased-MIMO radar is shown to offer a much better overall performance despite the small loss in performance that it shows in the transmit beampattern as compared to the phased-array radar, and the small loss in performance that it shows in the waveform diversity beampattern as compared to the MIMO radar.

Another observation drawn from Figs. 5–7 is that the spatial aliasing in the transmit and the waveform diversity beampatterns has an effect on the overall transmit/receive beampattern. As a result, the sidelobe levels for all radar techniques exhibit large variations over the spatial range. In particular, the sidelobe levels of the overall beampattern in the areas which correspond to the locations of the repeated mainlobe in the transmit and/or the waveform diversity beampatterns can be 30 dB higher than the sidelobe levels of the overall beampattern which correspond to the repeated sidelobe regions in the transmit beampatterns and/or the waveform diversity beampatterns. This phenomenon warrants the necessity for imposing some sort of sidelobe control on the design of transmit/receive beamforming for MIMO radar systems. However, due to space limitations, this opportunity is not considered here.

*Example 3: Non-adaptive output SINR*

In this example, the non-adaptive beamformer output SINR is tested versus SNR for different INR values. Fig. 8 shows the output SINR versus SNR (INR is fixed to 30 dB) for the phased-array, MIMO, and phased-MIMO radars. It can be seen from the figure that the output SINR for the phased-array and MIMO radars are almost the same. This observation agrees with the fact that both techniques have the same sidelobe attenuation level and, therefore, have the same interference rejection capabilities. On the other hand, the phased-MIMO radar has a much higher output SINR as compared to both the phased-array and MIMO radars.

Fig. 9 shows the output SINR versus SNR while the INR is fixed to $-30$ dB, i.e. the interference power can be neglected as compared to the noise power. It can be seen from the figure that the phased-array radar output SINR is 10 times higher than the MIMO radar output SINR. This gain is attributed to the transmit coherent processing gain that the phased-array radar enjoys. It is also observed from this figure that the phased-MIMO radar has an output SINR that is very close to the output SINR of the phased-array radar yet it enjoys the waveform diversity benefits offered by the MIMO radar. This depicts the tradeoff offered by the phased-MIMO radar.

*Example 4: Non-adaptive output SINR in the presence of spatially distributed interference* In this example, we assume one spatially distributed interference source which is uniformly spread over the spatial sector $[-50°, -20°]$. The spatial power density of the distributed source is normalized such that the total power of the interference is equivalent to the required INR. We assume that the target has the same power as the interference, i.e., INR=SNR. Fig. 10 shows the output SINR versus SNR where the INR is varied. It can be seen from the figure that at low SNR the phased-array radar output SINR is higher than the MIMO radar output SINR. However as the SNR increases, the difference between the SINRs of both techniques tend to decrease and the two curves eventually coincide. On the other hand, the proposed phased-MIMO radar is shown to have an output SINR that outperforms both techniques at SNR=INR larger than 10 dB. At low SNR values, the output SINR of the phased-MIMO radar is comparable to the output SINR of the phased-array radar which coincides with our theoretical founding.

## B. Adaptive Transmit/Receive Beamforming

*Example 5: MVDR beamforming employing multiple transmit multiple receive antennas*

In this example, the MVDR receive beamforming (57) is used for the phased-array, MIMO, and phased-MIMO radars. All simulation parameters are the same as in Example 1 except that the target power is fixed to 0 dB while the interference power is fixed to 50 dB. At the transmitting array, the conventional beamformer is used and, therefore, the transmit beampatterns for all radar techniques tested are the same as those given in Fig. 2. The receive MVDR beampattern is shown in Fig. 11 for all radar techniques tested. It can be observed from this figure that all radar techniques exhibit nulls at the locations of the powerful interference. Moreover, the proposed phased-MIMO radar as well as the phased-array radar have lower sidelobe levels as compared to the MIMO radar. This means that the phased-MIMO radar has almost the same robustness against sensor noise as the phased-array radar. At the same time, it enjoys the advantages of the MIMO radar, e.g., waveform diversity.

Fig. 12 shows the optimal SINR as well as the MVDR output SINR versus SNR (INR is fixed to 30 dB) for all radar techniques tested. From this figure, we can see that the phased-array radar outperforms the MIMO radar that can be attributed to robustness of the phased-array radar against sensor noise due to the use of transmit coherent processing. On the other hand, the phased-MIMO radar exhibits SINR performance that is very close to the phased-array SINR performance. It is interesting to note that the proposed phased-MIMO radar offers a substantially better MVDR output SINR as compared to the MIMO radar. This gain is expected since the proposed MIMO radar combines the advantages of both the phased-array and MIMO radars.

*Example 6: MVDR beamforming employing multiple transmit single receive antennas*

In the last example, we investigate an interesting case of a bistatic radar system which employs multiple transmit antennas and a single receive antenna. The same scenario as in Example 5 is considered here except that the number of receive antennas is $N = 1$. In this case, the dimensions of the received data by the phased-array, MIMO, and phased-MIMO radar systems will be $1 \times 1$, $M \times 1$, and $K \times 1$, respectively. The MVDR beamformer (57) is used for the MIMO and phased-MIMO radar techniques and the MVDR beampatterns are plotted in Fig. 13 for INR = 50 dB. We observe from this figure that the phased-array radar has failed to reject the powerful interference. In fact, the adaptive beamforming is not helpful with the phased-array radar because the dimension of the received data is just $1 \times 1$. The main beam exhibited by the phased-array radar is attributed to the transmit coherent processing only. Contrary to the phased-array radar, the MIMO and phased-MIMO radars lend themselves easily to adaptive techniques at the receiving end. The virtual extended array of larger dimensionality at the received end is the result of using waveform diversity at the transmitting array. As we can see from Fig. 13, both the MIMO and phased-MIMO radars exhibit excellent adaptive interference rejection capabilities. However, the MIMO radar suffers from poor performance because of its lack of robustness against sensor noise. On the other hand, the phased-MIMO radar enjoys much lower sidelobe levels and, therefore, higher robustness against sensor noise as compared to the MIMO radar. Indeed, Fig. 13 demonstrates that the proposed phased-MIMO radar enjoys the advantages of phased-array and MIMO radars and, therefore, is superior to both.

Fig. 14 shows the optimal SINRs as well as the MVDR output SINRs versus SNR (INR is fixed to 30 dB) for all radar techniques tested. From this figure, we can see that the phased-array radar performs very poorly due to its inability to reject strong interference. On the other hand, the MIMO radar is

August 15, 2009　　　　　　　　　　　　　　　　　　　　　　　　　　　　　　　　　　　　　　　　　　DRAFT

able to adaptively reject interference and, therefore, it has better SINR performance as compared to the phased-array radar. Moreover, the phased-MIMO radar exhibits the SINR performance that is superior to the performance of both the phased-array and MIMO radars. This gain is attributed to the ability of the phased-MIMO radar to reject interference combined with its robustness against sensor noise.

## VI. CONCLUSIONS

The new technique for MIMO radar with colocated antennas has been proposed. This technique is based on partitioning the transmitting array into a number of subarrays which are allowed to overlap. Each subarray is used to coherently transmit a waveform which is orthogonal to the waveforms transmitted by other subarrays. Coherent processing gain is achieved by designing the weight vector of each subarray to form a beam towards a certain direction in space. The subarrays are combined jointly to form a MIMO radar resulting in higher resolution capabilities. It is shown that the proposed technique combines the advantages of the phased-array radar and the advantages of the MIMO radar and, therefore, it has a superior performance. Simulation results confirm our theoretical observations and demonstrate the effectiveness of the proposed phased-MIMO radar technique. The formulation of the new phased-MIMO radar technique opens a new avenue in MIMO radar developments.

## APPENDIX A: PROOF OF PROPOSITION 2

In this appendix, we prove that the highest sidelobe of the phased-MIMO radar beampattern is lower than the highest sidelobe of the phased-array radar beampattern, i.e., we prove that

$$\max_{\theta \in \bar{\Theta}} \frac{\left|\mathbf{a}_K^H(\theta_s)\mathbf{a}_K(\theta)\right|^2}{(M-K+1)^2} \cdot \frac{\left|\mathbf{d}^H(\theta_s)\mathbf{d}(\theta)\right|^2}{K^2} \leq \max_{\theta \in \bar{\Theta}} \frac{\left|\mathbf{a}^H(\theta_s)\mathbf{a}(\theta)\right|^2}{M^2}. \tag{58}$$

Using Fourier transform analysis, the square-root of the right hand side of (58) can be expressed as

$$\frac{\left|\mathbf{a}^H(\theta_s)\mathbf{a}(\theta)\right|}{M} = \frac{\delta(\Omega - \Omega_s)}{M} * |\mathtt{sinc}(M\Omega)|, \quad -\pi \leq \Omega \triangleq 2\pi\frac{d}{\lambda}\sin(\theta) \leq \pi \tag{59}$$

where $\delta(\Omega)$ is the Dirac-delta function, $d$ is the interelement spacing, $\lambda$ is the wavelength of the propagating wave, $*$ is the convolution operator, and $\mathtt{sinc}(\kappa\Omega)$ is given by

$$\mathtt{sinc}(\kappa\Omega) \triangleq \frac{\sin(\kappa\frac{\Omega}{2})}{\sin(\frac{\Omega}{2})} \tag{60}$$

for some positive integer $\kappa$. Similarly, the square-root of the left hand side of (58) can be expressed as

$$\frac{\left|\mathbf{a}_K^H(\theta_s)\mathbf{a}_K(\theta)\right|}{(M-K+1)} \cdot \frac{\left|\mathbf{d}^H(\theta_s)\mathbf{d}(\theta)\right|}{K} = \frac{\delta(\Omega - \Omega_s)}{K(M-K+1)} * (|\mathtt{sinc}((M-K+1)\Omega)| \cdot |\mathtt{sinc}(K\Omega)|)$$
$$= \delta(\Omega - \Omega_s) * H_K(\Omega) \tag{61}$$

where

$$H_K(\Omega) = \frac{(|\texttt{sinc}((M-K+1)\Omega)| \cdot |\texttt{sinc}(K\Omega)|)}{K(M-K+1)}. \tag{62}$$

The function $H_K(\Omega)$ is plotted in Fig. 15 for different values of $K$. It can be observed from the figure that the case $K = 1$ which corresponds to the phased-array radar has higher sidelobes as compared to all other cases when $K > 1$.

Let us now state some properties related to the function $\texttt{sinc}(\kappa\Omega)$.

P1) The main lobe occupies the region $0 \leq |\Omega| \leq 2\pi/\kappa$.

P2) For $\kappa = 2$, the main lobe occupies the whole region $0 \leq |\Omega| \leq \pi$ while side lobes do not exist.

P3) For $\kappa = 3$, there is only one sidelobe whose peak is located at $\Omega = \pi$; at this point $\sin(\Omega/2) = 1$.

P4) For $\kappa \geq 4$, there are multiple sidelobes. The sidelobe closest to the main lobe occupies the region $2\pi/\kappa \leq |\Omega| \leq 4\pi/\kappa$.

P5) Noting that $\sin(\Omega/2)$ is monotonically increasing within $[0, \pi]$, the highest peak sidelobe is the one closest to the main lobe.

P6) The peak of the highest sidelobe occurs when $\sin \kappa\Omega/2$ is approximately at its maximum, i.e, $\sin \kappa\Omega/2 \cong 1$ which occurs at $\Omega \cong 3\pi/\kappa$ [27, Ch. 2].

Let $\zeta_1$, $\zeta_2$, and $\zeta_3$ be the peaks of the highest sidelobes of $\texttt{sinc}(M\Omega)/M$, $\texttt{sinc}((M-K+1)\Omega)/(M-K+1)$, and $\texttt{sinc}(K\Omega)/K$, respectively. Let also $\Omega_1$, $\Omega_2$, and $\Omega_3$ be the locations of $\zeta_1$, $\zeta_2$, and $\zeta_3$, respectively. Then, the maximum sidelobe of the left hand side of (58) is upper-bounded as follows

$$\max_{\Omega} H_K(\Omega) \leq \zeta_2\zeta_3. \tag{63}$$

Therefore, the inequality (58) can be proved by proving that

$$\zeta_2\zeta_3 < \zeta_1. \tag{64}$$

Dividing $\zeta_2\zeta_3$ by $\zeta_1$, we obtain the ratio

$$\Gamma \triangleq \frac{\zeta_2\zeta_3}{\zeta_1} = \frac{\sin\left((M-K+1)\frac{\Omega_2}{2}\right)}{(M-K+1)\sin\left(\frac{\Omega_2}{2}\right)} \cdot \frac{\sin\left(K\frac{\Omega_3}{2}\right)}{K\sin\left(\frac{\Omega_3}{2}\right)} \cdot \frac{M\sin\left(\frac{\Omega_1}{2}\right)}{\sin\left(M\frac{\Omega_1}{2}\right)}$$
$$= \underbrace{\frac{M}{K(M-K+1)}}_{\alpha_1} \cdot \underbrace{\frac{\sin\left(\frac{\Omega_1}{2}\right)}{\sin\left(\frac{\Omega_2}{2}\right)\sin\left(\frac{\Omega_3}{2}\right)}}_{\alpha_2} \cdot \underbrace{\frac{\sin\left((M-K+1)\frac{\Omega_2}{2}\right)\sin\left(K\frac{\Omega_3}{2}\right)}{\sin\left(M\frac{\Omega_1}{2}\right)}}_{\alpha_3}. \tag{65}$$

Using Property (P6), $\alpha_3$ can be approximated as

$$\alpha_3 \cong \sin\left((M-K+1)\frac{\Omega_2}{2}\right)\sin\left(K\frac{\Omega_3}{2}\right) \leq 1. \tag{66}$$

Moreover, noting that $K(M - K + 1) - M = (M - K)(K - 1) \geq 0$, we conclude that $\alpha_1 \leq 1$. Using Properties (P4–P6), we find that $\Omega_1 \cong 3\pi/M$. Therefore, $\sin(\Omega_1/2)$ can be upper-bounded as follows

$$\sin\left(\frac{\Omega_1}{2}\right) < \frac{\Omega_1}{2} \cong \frac{3\pi}{2M}. \tag{67}$$

On the other hand, we have $\Omega_2 \cong 3\pi/(M - K + 1)$. Moreover, using Property (P4) we conclude that $\Omega_2 \in [2\pi/(M - K + 1),\ 4\pi/(M - K + 1)]$. Therefore, $\sin(\Omega_2/2)$ can be lower-bounded as follows

$$\sin\left(\frac{\Omega_2}{2}\right) \geq \frac{1}{2} \cdot \frac{2\pi}{M - K + 1} = \frac{\pi}{M - K + 1}, \quad M - K + 1 > 3. \tag{68}$$

Similarly, using the same chain of arguments as above, $\sin(\Omega_3/2)$ can be lower-bounded as follows

$$\sin\left(\frac{\Omega_3}{2}\right) \geq \frac{\pi}{K}, \quad K > 3. \tag{69}$$

Therefore, using (67), (68), and (69), we obtain

$$\alpha_2 = \frac{\sin\left(\frac{\Omega_1}{2}\right)}{\sin\left(\frac{\Omega_2}{2}\right)\sin\left(\frac{\Omega_3}{2}\right)} \leq \frac{\frac{3\pi}{2M}}{\frac{\pi}{K}\frac{\pi}{M-K+1}} = \frac{3}{2\pi\alpha_1}. \tag{70}$$

Substituting (66) and (70) in (65), we prove that $\Gamma \leq 3/(2\pi) < 1$. This means that $\zeta_2\zeta_3 < \zeta_1$ which proves (58) for the case when $K > 3$ and $M - K + 1 > 3$. Using property (P3), for the case when $K = M - K + 1 = 3$, we have $\sin(\Omega_2/2) = \sin(\Omega_3/2) = 1$. Hence, in this case, we have $\alpha_2 < 3\pi/2M < 1$ yielding $\Gamma < 1$. Therefore, (58) is fully proved. ∎

## ACKNOWLEDGMENT

The authors would like to thank Dr. Y. Abramovich of Australian Defence Science and Technology Organisation (DSTO), Adelaide, for helpful discussions and clarifications on power normalization for the output SINR of the phased-MIMO radar and Dr. A. Gershman of Darmstadt University of Technology for useful discussions.

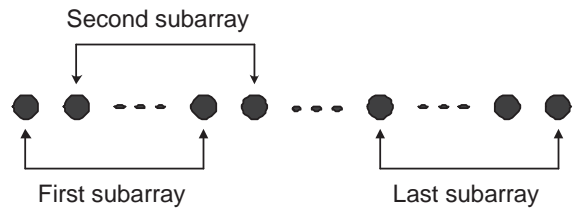

Fig. 1.  Transmit array of $M$ antennas partitioned into $K$ subarray of $M-K+1$ antennas each.

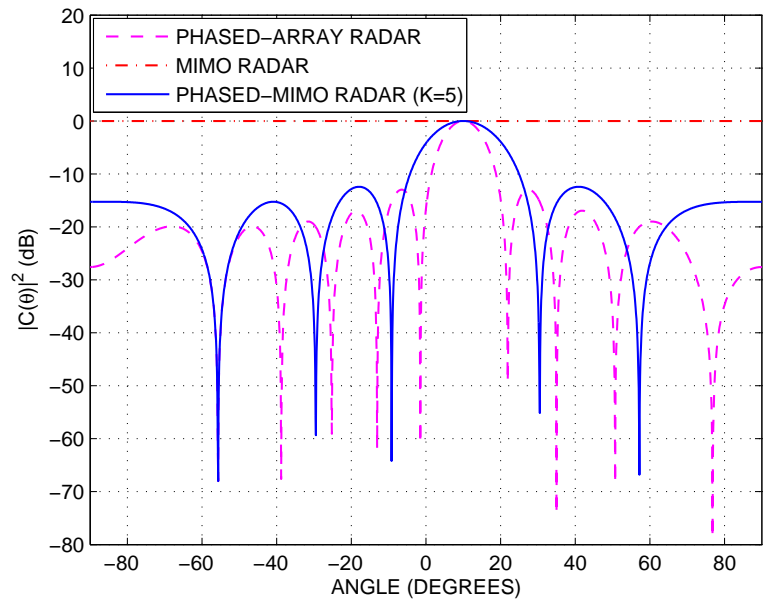

Fig. 2.  First Example: Transmit beampatterns using conventional beamformer ($d_T = 0.5$ wavelength).

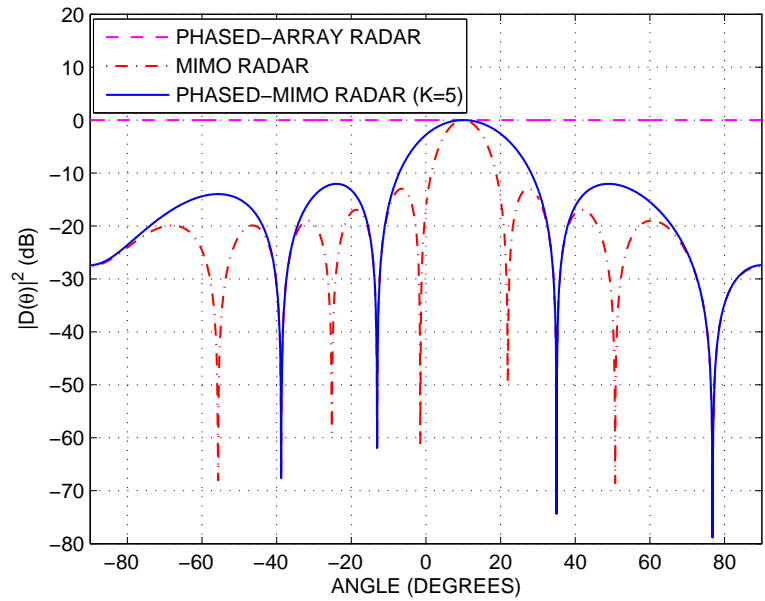

Fig. 3. First Example: Waveform diversity beampatterns using conventional beamformer ($d_T = 0.5$ wavelength).

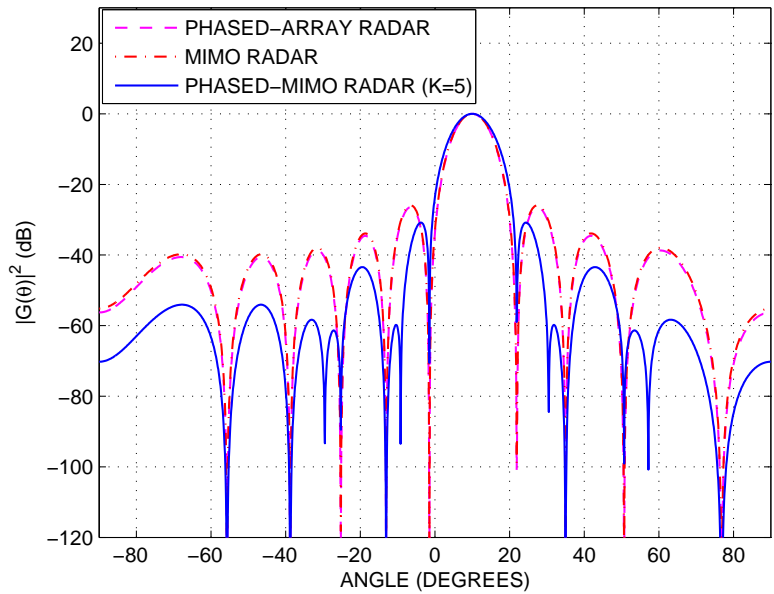

Fig. 4. First Example: Overall beampatterns using conventional transmit/receive beamformer ($d_T = 0.5$ wavelength).

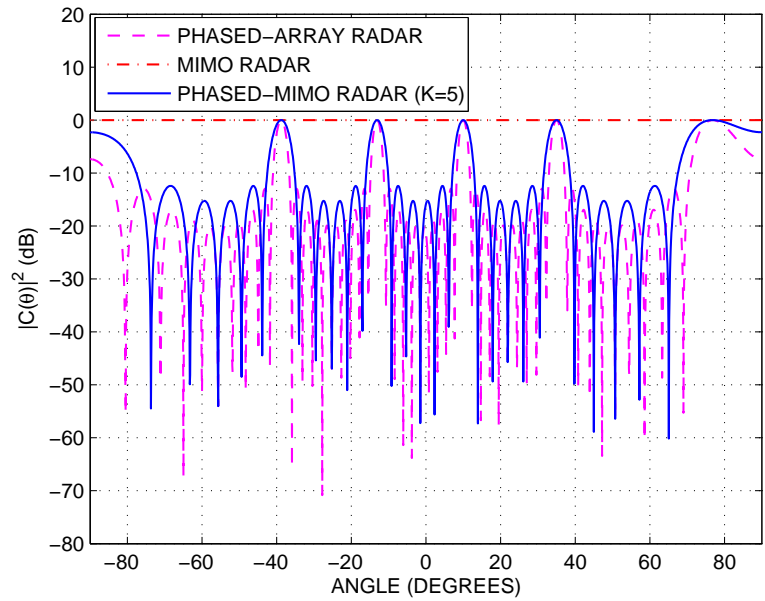

Fig. 5. Second example: Transmit beampatterns using conventional beamformer ($d_T = 2.5$ wavelength).

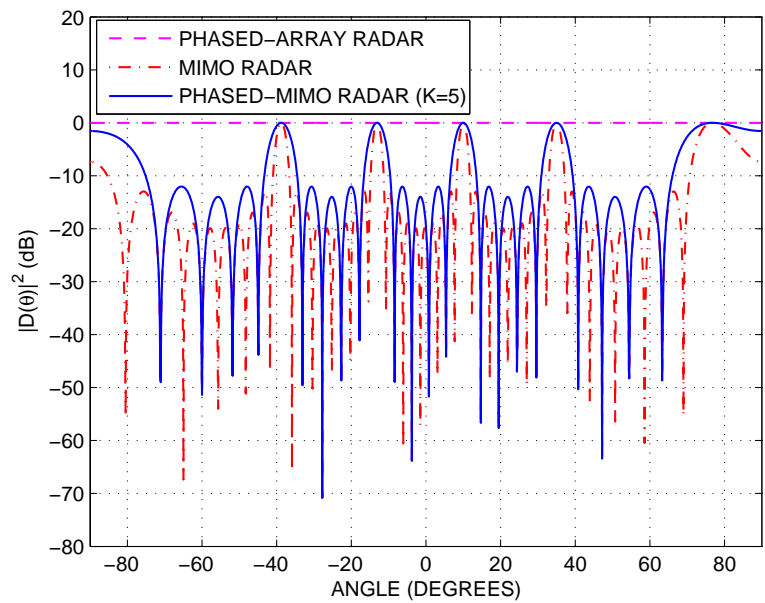

Fig. 6. Second example: Waveform diversity beampatterns using conventional beamformer ($d_T = 2.5$ wavelength).

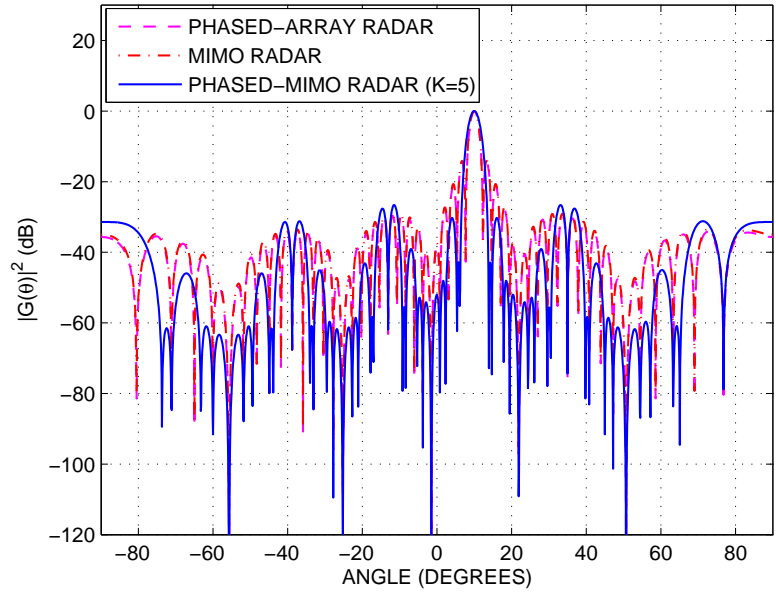

Fig. 7. Second example: Overall beampatterns using conventional transmit/receive beamformer ($d_T = 2.5$ wavelength).

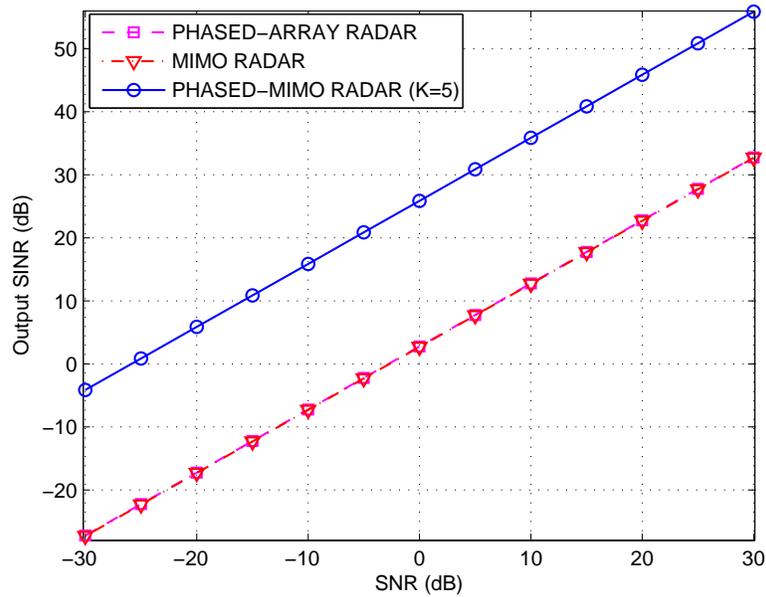

Fig. 8. Third example: Non-adaptive transmit/receive output SINRs versus SNR at fixed INR=30 dB.

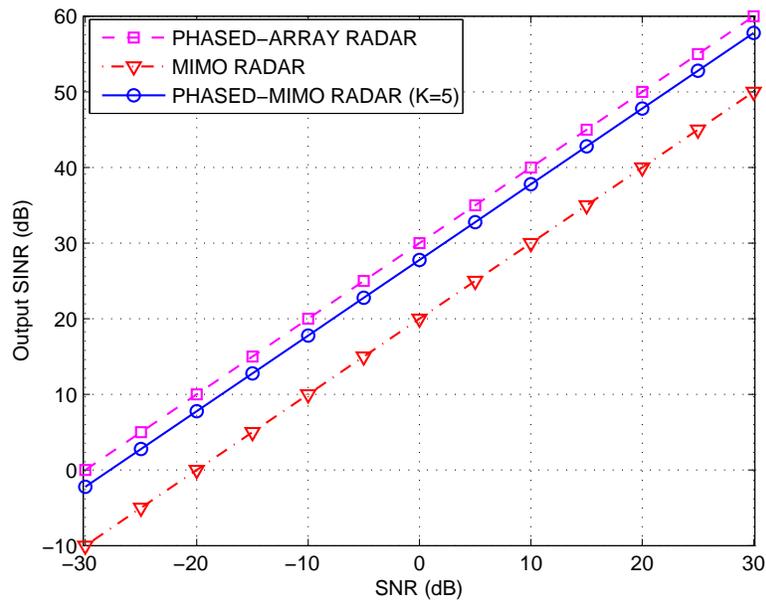

Fig. 9. Third example: Non-adaptive transmit/receive output SINRs versus SNR at fixed INR=−30 dB.

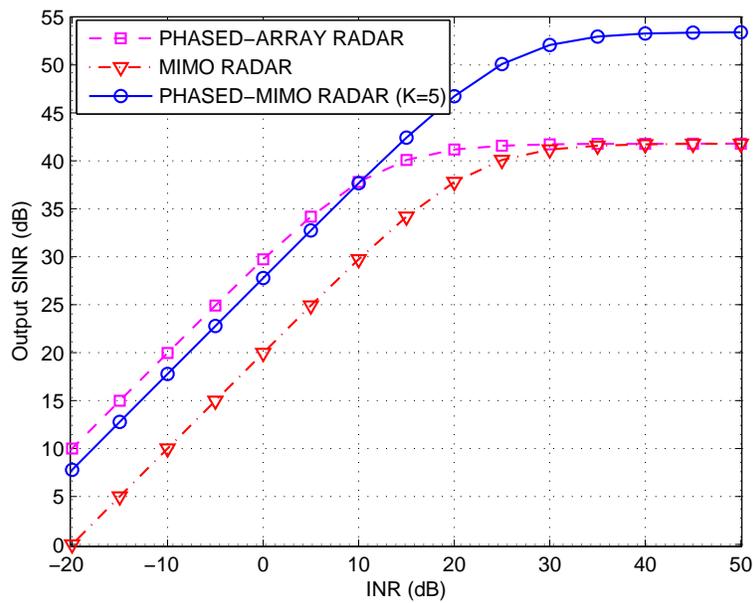

Fig. 10. Fourth example: Non-adaptive transmit/receive output SINRs versus INR=SNR; spatially distributed interfernce.

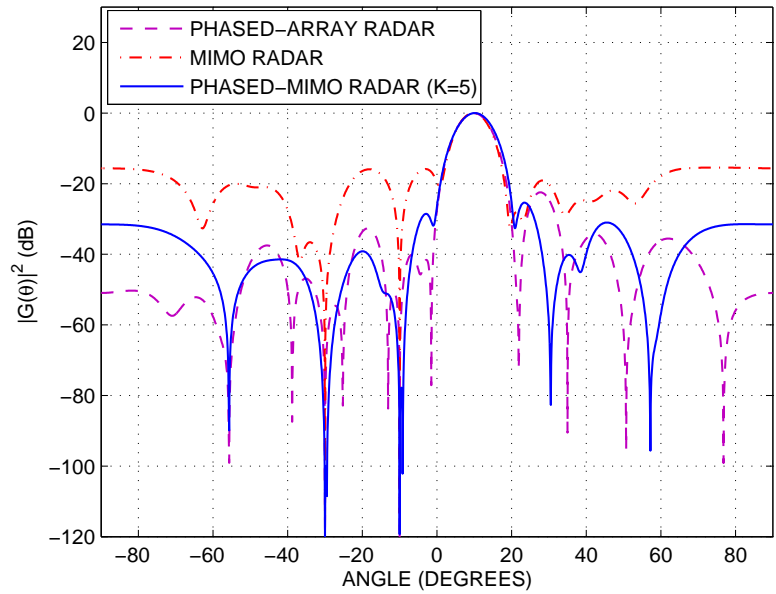

Fig. 11. Fifth Example: Overall beampatterns using MVDR beamformer ($d_T = 0.5$ wavelength).

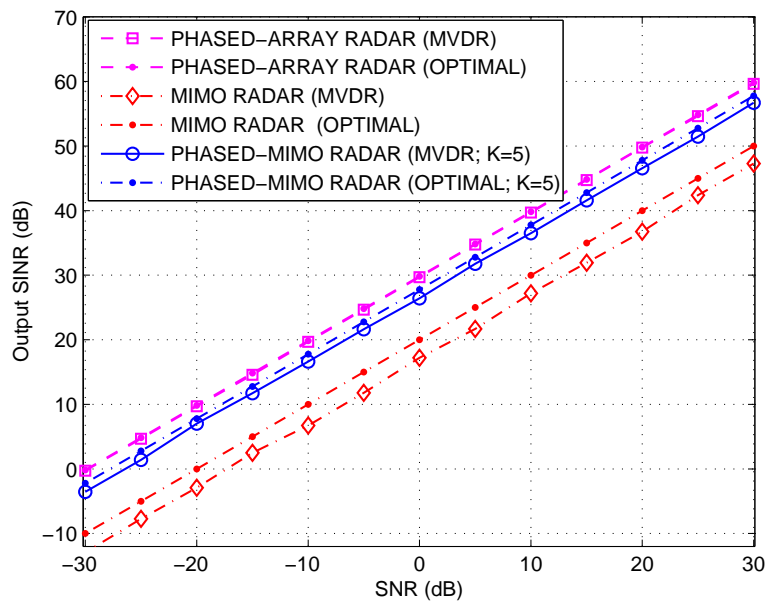

Fig. 12. Fifth Example: Output SINRs versus SNR, $N = 10$ receiving antennas spaced half wavelength apart.

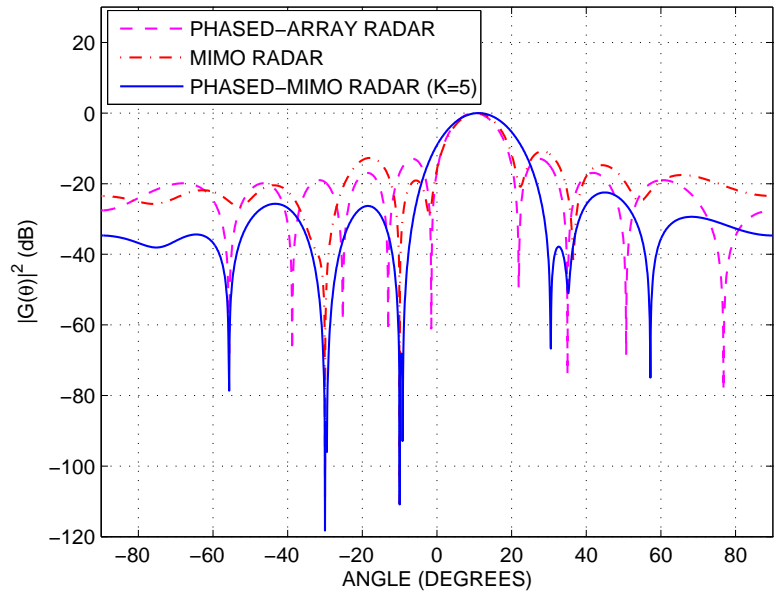

Fig. 13. Sixth Example: Overall beampatterns using MVDR beamformer, $N = 1$ receiving antenna.

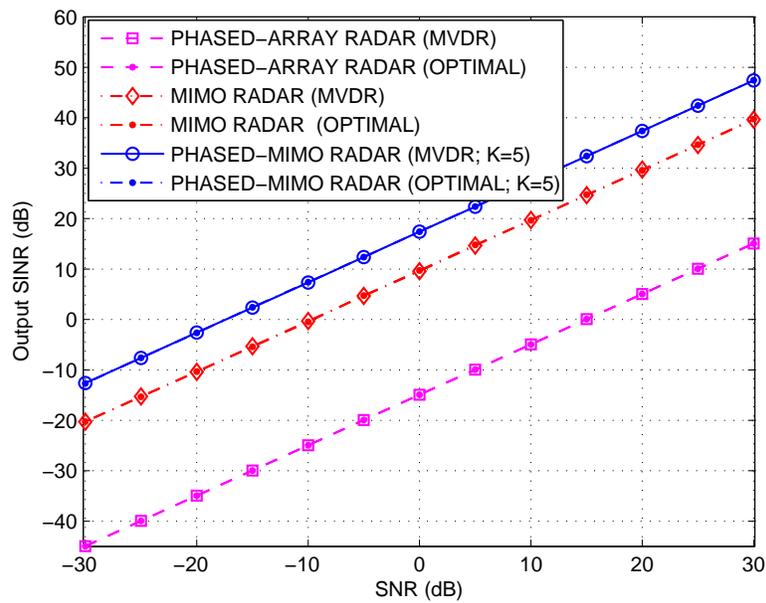

Fig. 14. Sixth Example: Output SINRs versus SNR, $N = 1$ receiving antenna.

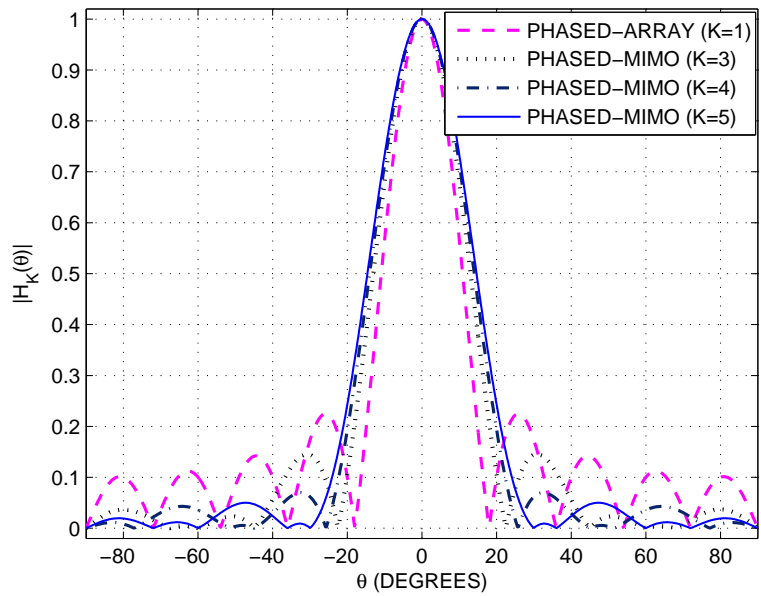

Fig. 15. Comparison between sidelobe levels of the phased-MIMO and phased-array radars (number of transmit antennas is fixed to $M = 10$); $H_K(\theta) = \frac{|\text{sinc}((M-K+1)\Omega)| \cdot |\text{sinc}(K\Omega)|}{K(M-K+1)}\Big|_{\Omega=2\pi\frac{d}{\lambda}\sin(\theta)}$.